\documentclass[aip,imp,amsmath,amssymb,reprint]{revtex4-1}

\draft 
\usepackage{graphicx}
\usepackage{dcolumn}

\begin{document}


\title{Expectations from the Liouville--von Neumann Equation Using Chebyshev Expansion} 



\author{Giacomo Mazzi}
\email[Electronic Mail: ]{G.Mazzi@sms.ed.ac.uk}
\author{Benedict J. Leimkuhler}
\affiliation{School of Mathematics, University of Edinburgh, Edinburgh EH9 3JZ, UK}

\date{\today}

\begin{abstract}
We consider a natural dimension reduction technique for the Liouville-von Neumann equation for a mixed quantum system based on evaluation of a trace formula combined with a direct expansion in modified Chebyshev polynomials.   This reduction is highly efficient and does not destroy any information.  We demonstrate the practical application of the scheme with a model problem and compare with popular alternatives.  This method can be applied to autonomous quantum problems where the desired outcome of quantum simulation is the expectation of an observable.
\end{abstract}


\maketitle 

\section{The Liouville Von--Neumann Equation}
An ensemble of $N$ quantum systems each described by a wave function $|\Psi\rangle$ may be expressed in terms of a density operator $\varrho$ defined by:
\begin{equation}
\varrho=\sum_{j=1}^N |\Psi^j\rangle\langle\Psi^j|.
\end{equation}
The temporal evolution of this quantity is characterized by the Liouville-von Neumann equation:
\begin{equation}\label{liovvn}
 \frac{\partial \varrho}{\partial t}=-i[H,\varrho], 
\end{equation}
where we have set $\hbar=1$. If $\varrho$ is expanded using a (finite) approximate basis set $\{|\varphi_1\rangle\ldots |\varphi_n\rangle\}$,  (\ref{liovvn}) may be viewed as an ordinary differential equation in a matrix argument.
In the time independent case the solution for (\ref{liovvn}) may be written:
\begin{equation}\label{sol}
 \varrho(t)=e^{-iH t}\varrho_0 e^{iH t}.
\end{equation}

It is possible and sometimes preferable to rewrite (\ref{sol}) by introduction of the 
Liouvillian $L= \rm{Id} \otimes H-H \otimes \rm{Id}$, where $\rm{Id}$ is the identity matrix,  allowing us to recast $\varrho$ as a vector:
\begin{equation}\label{liov}
\varrho(t)=e^{-iL t}\varrho_0.
\end{equation}
The main issue is then to evaluate the exponential of the matrix $L$.
In the typical case, the size of $\varrho$ grows exponentially in the number of particles; 
at the same time, for a large system and using a typical basis set, $H$ (and $L$) will be a structured sparse matrix.

In the past decades many methods have been developed for numerical evaluation of the matrix exponential \cite{nineteen}.
However for large sparse matrices, the methods usually applied are those based on expansion in Krylov subspace \cite{saad1,lubich1,schulze}.
The main idea is to project (\ref{liov}) onto the subspace:
\begin{equation}\label{kry_sub} 
K_m(L,\varrho)={\rm span} \{\varrho_0,L\varrho_0,L^2\varrho_0,\ldots,L^m\varrho_0\}.
\end{equation}
To get a suitable basis set of (\ref{kry_sub}) we may use the Lanczos algorithm \cite{lancz}, as  $L$ is Hermitian.
The Lanczos method is an iterative method, a very desirable feature in the context of large sparse matrices.
For specific details see Ref.~\onlinecite{golubok} and Ref.~\onlinecite{cullum}.
However, because of the fact that all these methods involve the propagation of the matrix $\varrho$, 
they suffer from requiring that matrix operations (or matrix-vector operations) be performed at each step of calculation.

While the evolution of the system is described by the density matrix, the outputs we are interested in obtaining from quantum simulations are
the expectations of observables, these being the only quantities we can compare with the experiments.
In the density matrix formalism the expectation value of an observable $Q$, associated with an operator $\hat{Q}$ is written as:
\begin{equation}\label{obs}
 \langle \hat{Q}(t) \rangle={\rm Trace} \{\varrho(t) \hat{Q}\}.
\end{equation}
Whereas in general quantum simulations the equation of motion is solved for a quantity, $\varrho$, which has dimension $n \times n$, 
the types of outputs we are generally interested in are just one dimensional objects (\ref{obs}).
In this paper we exploit this fact and design an algorithm that computes (almost) directly the evolution of the expectation value (\ref{obs}),
instead of the evolution of the density matrix (\ref{sol}).
This approach does not lose any information of the original system (the only errors arise due to truncation), but at the same time the method
provides a powerful computational tool,  with potential dramatic reduction in the computational cost, especially when dealing with large matrices.
The main idea of this approach is to exploit features of a Chebyshev expansion for the matrix exponential in (\ref{liov}). 
It is important to remark that Chebyshev polynomials have been widely used to get polynomial approximations of function of matrices, and especially for the matrix exponential in (\ref{sol}) \cite{tal}. The terms of the Chebyshev expansion can be constructed
iteratively with the price being a sequence of matrix--matrix multiplications.

Several methods to evaluate (\ref{liov}) are discussed in a recent monograph, see Ref.~\onlinecite{lubichbok}; however our  proposed {\em Direct evaluation of the Expectation values via Chebyshev polynomial} (DEC) method  is different because it does not solve the evolution for the density matrix. Instead it exploits the trace evaluation in (\ref{obs}) and with the evaluation of just one Chebyshev expansion it allows the solution of (\ref{obs}) at any time. 
DEC can be extremely powerful when we are only interested in the expectation values; if instead it is necessary to evaluate the evolution of the density matrix itself (\ref{liov}), then a traditional approach might be the best choice.

\section{The Direct Expectation values via Chebyshev}
The preliminary step of our method is to rescale the matrix within the interval $[-1,1]$, as outside this interval the Chebyshev polynomials grow rapidly, and the expansion becomes unstable; to do that we need to evaluate the two extremes of the spectrum of $L$.   
In order to obtain extreme values we propose, as already mentioned in the literature \cite{saad_lanc}, to perform a few steps of Lanczos iteration, as this provides a good approximation for the extreme eigenvalues, for small computational cost. 
If we define these two values as $\alpha$ and $\beta$, i.e. $\beta\leq \sigma(L)\leq \alpha$, we may rewrite $L$ as $L=(S\,{\rm Id}-L_sD)$, where $D=(\alpha-\beta)/2$, $S=(\alpha+\beta)/2$, and $-1\leq\sigma(L_s)\leq 1$. We may then expand the exponential of $L_s$ in the Chebyshev polynomials and we
arrive at the following equation for $\varrho$:
\begin{equation}\label{chebexp}
\varrho(t)=e^{-iLt}\varrho_0\approx e^{-itS}\left(\sum_{k=0}^{n_{\rm max}}c_k(t_D) T_k(L_s)\varrho_0\right),
\end{equation}
with $t_D=Dt$.
Both $c_k(t_D)$ and $T_k(L_s)$ can be calculated iteratively:
\begin{subequations}\label{cktk}
\begin{equation}\label{ck}
 c_k(t)=(2-\delta_{k,0})(-i)^k J_k(t),
 \end{equation}
 \begin{equation}\label{Tk}
T_{k+1}(x)=2T_k(x)x-T_{k-1}(x),
\end{equation}
\end{subequations}
with initial values $T_0(x)={\rm Id}$, $T_1(x)=x$.
$J_k(t)$ is the $k$-th Bessel function of the first kind.

If we insert (\ref{chebexp}) into (\ref{obs}) we find:
\begin{equation}\label{obs_cheb}
\langle \hat{Q}(t)\rangle={\rm Trace}\left\{ \left(e^{-i t S}\sum_{k=0}^{n_{\rm max}}c_k(t_D) T_k(L_s)\varrho_0\right) \hat{Q}\right\}.
\end{equation}
By exploiting the linearity of the trace operation we can pull out of the trace all time dependent parts, and evaluate once 
for all the coefficients $T_k(L_s)$.
In fact we may rewrite  (\ref{obs_cheb}) as:
\begin{equation}\label{obs_cheb2}
 \langle \hat{Q}(t)\rangle=e^{-itS}\sum_{k=0}^{n_{\rm max}}c_k(t_D) {\rm Trace}\left\{ T_k(L_s)\hat{Q}\right\}.
\end{equation}
This is the key equation of the DEC method as it is possible to store an array of scalar values $\tilde{T}_k={\rm Trace}\{(T_k\varrho_0)\hat{Q}\}$. All the time dependent terms are just scalar values that have to be multiplied by $\tilde{T}_k$ to get the evolution of $\hat{Q}$ at any time:
\begin{equation}\label{obs_cheb3}
 \langle \hat{Q}(t)\rangle=e^{-itS}\sum_{k=0}^{n_{\rm max}}c_k(tD) \tilde{T}_k.
\end{equation}

If more than one observable is required it is still possible to use DEC. The only difference with the single expectation case is that we need to store different sets of $T_k$, one for each operator $\hat{Q}$.

\subsection{Stopping Criterion}
The number of terms for the polynomial expansion in (\ref{chebexp}) depends on a prescribed tolerance $\varepsilon$, and on the time $t_D$. In other methods based on Chebyshev approximation \cite{spinev}, the following has been suggested as a stopping criterion:
\begin{equation}\label{c_max}
n_{\rm max}\quad {\rm s.t.}\quad \|c_{n_{\rm max}}(t_D)\|<\varepsilon.
\end{equation}
Due to the zeros of the Bessel function $J$, at fixed time $t_D$, (\ref{c_max}) may hold for some $n$, even though the expansion has not yet reached the convergence regime; it may happen that for $n_1>n$ we have that $c_{n_1}(t_D)>c_n(t_D)$.
To avoid this effect it is enough to use as a stopping criterion a combination of two Bessel functions; the cost of such a stopping criterion is that at most we need to perform an extra iteration step (\ref{ck}).
In our numerical tests we have used the following:
\begin{equation}\label{c_max1}
 n_{\rm max}\quad {\rm s.t.}\quad \sqrt{\|c_{n_{\rm max}-1}(t_D)\|^2+\|c_{n_{\rm max}}(t_D)^2\|}<\varepsilon.
\end{equation}
The total time $\tau$ plays a role here, since the larger $\tau$ the more terms $(T_k,c_k)$ will be needed to get $| c_k|$ below
 the threshold $\varepsilon$.

\subsection{Computation of the Expansion}
In order to optimise the number of terms we evaluate, but without having to check at each step whether we have already evaluated enough terms $T_k$,  we propose to evaluate first $\langle \hat{Q}(t)\rangle$, at the final time $\tau$, and to store the $N_{\rm max}$ values of $\tilde{T}_k$.

From equation (\ref{ck}) it is clear that $c_k$ depends on the Bessel functions.  If we look at the asymptotic behavior of the Bessel function of fist kind, for any $k\in N$, we have that, for $k$ fixed \cite{handbok}:
\begin{equation}\label{asympt}
J_k(t)\sim \frac{1}{\Gamma(k+1)}\left(\frac{t}{2}\right)^k, \qquad  \lim t\rightarrow 0,
\end{equation}
where $\Gamma(t)$ is the Euler--$\Gamma$ and for $n\in Z$ we have that $\Gamma(n)=(n-1)!$.
Equation (\ref{asympt}) shows that for any $k\neq 0$, in a neighbourhood of $t=0$, $J_k(t)$ is increasing monotonically with respect to $t$.
This behavior is maintained for the whole interval $[0,j_k']$ where $j_k'$ is the first zero of the derivative of $J_k$.
It is possible to show (see Ref.~\onlinecite{handbok},  Eq.$9.5.2$), that $k\leq j_k'$;  consequently we can say that if (\ref{c_max1}) holds for a given $n_{\rm max}$ at $\tau$ and $\tau\leq n_{\rm max}$, then we are in the monotonically increasing region for $J_{n_{\rm max}}$ and $J_{n_{\rm max}+1}$.  In this case, equation (\ref{c_max1}) holds also for any $t\leq \tau$.

\begin{figure}
\includegraphics[width=0.45\textwidth]{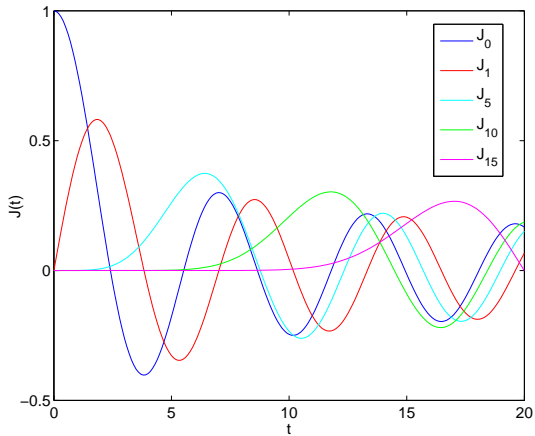}
\caption{Example of few integer order Bessel Functions of the first kind. } 
\label{fig:bessel}
\end{figure} 

The Bessel Functions of the First Kind  of integer order may be evaluated directly by using a three-term recurrence relation
\begin{equation}\label{besselj}
J_{n+1}(t)=\frac{2n}{t}J_n-J_{n-1}.
\end{equation}
It is well known that  (\ref{besselj}) becomes numerically unstable for $n>t$, see Ref.\onlinecite{backward}.
To improve the method, we may exploit the linear nature of the iterative algorithm. It is possible to use  Miller's algorithm, and to solve an inverted form of (\ref{besselj}), i.e. to solve for $J_{n-1}$ given $J_n$, $J_{n+1}$ \cite{backward}. When using  Miller's Algorithm it is suggested to expand the number of  terms (providing a sort of buffer), i.e. to start the backward iteration process from $m_{\rm start}=n+r$, where $n$ is the actual order of the function we are interested on and $r$ is some small expansion.  In this case we need to know already from an a priori error analysis how many iterations need to be performed to to get below the threshold $\varepsilon$.

It is possible to prove that for the rescaled Hermitian matrix $L_s$, when applied to a vector of unit Euclidian norm we have \cite{lubichbok}:
\begin{equation}\label{errlub}
\|P_{m-1}(tL_s)\varrho_0-e^{-it L_s}\varrho_0\|\leq 4(e^{1-(t/2m)^2}\frac{t}{2m})^m \, {\rm for}\, m>t
\end{equation}
where $P_m(t)$ is the order $m$ expansion in Chebyshev polynomials.
This equation indicates that there is a superlinear decay of the error when $m> t$.  The formula may be derived by examining the asymptotic behavior of the Bessel functions (\ref{asympt}).
We may then use the relation $4(\exp\{1-(\tau/2m)^2\}\frac{\tau}{2m})^m\leq \varepsilon$ to approximate $m$.

\subsection{Efficient Implementation}
The cost of DEC is all in the first step.   Note that the cost of the evaluation of any $\tilde{T}_k$ itself is roughly equivalent to that of 
a matrix--matrix multiplication, as per the iteration $T_{k+1}(L_s)=2T_k(L_s)-T_{k-1}$ (\ref{Tk}). But what is actually needed in all our calculations is $T_k\varrho_0$. Because of the linearity of the iterative expression, we may multiply $T_0$ and $T_1$ by $\varrho_0$ and then use (\ref{Tk}) directly on $T_k\varrho_0$. The iterated operation in then just a matrix--vector multiplication.

After all the $\tilde{T}_k$ needed have been stored, it is possible to get $\langle Q(t)\rangle$ at any time $t\in[0,\tau]$ by evaluating:
\begin{equation}\label{Qt}
\langle Q(t)\rangle=e^{-i Dt_s} \sum_{k=1}^{m_{max}} c_k(t_s) \tilde{T}_k
\end{equation} 
where the $c_k$ are evaluated iteratively via (\ref{ck}), and $m_{max}$ satisfies (\ref{c_max1}) for $tD$.\\
For the details of the algorithm we refer to the Appendix.

\section{Numerical Experiments}

Nuclear Magnetic Resonance (NMR) is a spectroscopy technique that exploits the interaction between nuclear spins and electromagnetic fields in order to analyse the samples. The temporal evolution of such a system is described via a density matrix that has size $(2I+1/2)^n$ where $I$ is the spin and $n$ the number of nuclei. The exponential growth of the size of $\varrho$ with respect to $n$ impedes the use of simulations when dealing with systems involving more than few ($5$-$6$) spins. Many attempts have been made to solve this (see e.g. Ref.\onlinecite{simpson}, Ref.\onlinecite{kuprov2} for recent approaches), even using Chebyshev polynomials \cite{spinev}.
These algorithms have been developed to simulate both liquid systems, where the Hamiltonian is generally time independent, and for Solid--State NMR.
In the last case the Hamiltonian is time dependent due to the non averaging out of  anisotropic interactions during the motion of the sample.
When the Hamiltonian is time dependent it is not possible to apply DEC, as it is not possible anymore to isolate the time dependent part out of the trace. 

As in many other physical systems, nuclear spin dynamics provides a perfect example to test DEC, because the final outcome of the simulations is an observable, the free induction decay (FID) signal, and this result is the sole important quantity, as it is the only data available from experiment.

We have used DEC to evaluate this quantity:
\begin{equation}\label{fid}
f(t)={\rm Trace } \left\{\varrho(t) I_p\right\}
\end{equation}
$\varrho(t)$ can be written as combination of Pauli matrices, and $I_p$ is the shift up operator: $I_p=I_x+iI_y$. 
As Hamiltonian we assumed a sum of isotropic chemical shift and the isotropic term of a pair interaction called Homonuclear  J--couplings, that depends on the inner product $I_j\cdot I_k$ \cite{Levitt}:
\begin{equation}
  H=-\sum_{j=1}^n \omega_j I_j^z+\sum_{j,l=1}^n J_{jl} I_j\cdot I_l
\end{equation}
For the initial density matrix we set $\varrho_0= -I_y$, that is the result of the application of a so called  $x$-pulse to a sample already under the effect of a strong constant magnetic field along the $z$ direction \cite{Levitt}. This is the usual initial condition when the acquisition of  the signal starts.

\begin{figure}
\includegraphics[width=0.45\textwidth]{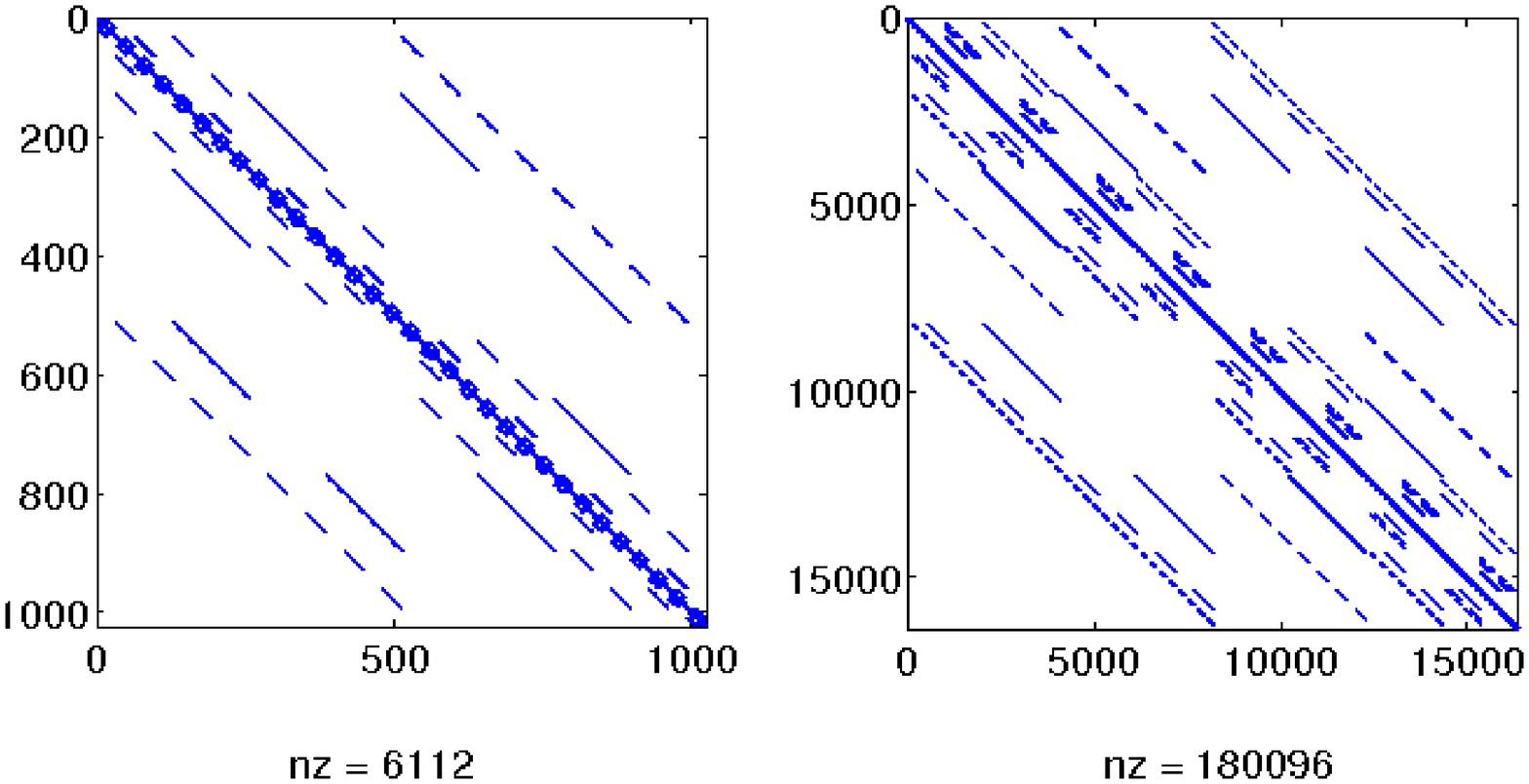}
\caption{Left: Sparse structure of the Liouvillian ($n=1024$, $nz=6112$) for a system of $5$ spins. Right: Structure of the Liouvillian ($n=16384$, $nz=180096$) for a system of $7$ spins. Approximately each spin is interacting with half the other spins in both cases.}
\label{fig:spy}
\end{figure}

An illustration of the structure of the Liouvillian matrix is presented in Fig.\ref{fig:spy}. The sparsity depends on the number of interactions among the spins. In most cases the J--coupling interaction matrix $J$ is relatively sparse. In our numerical simulations each spin interacts with about half of the other spins.

Due to the fact that our implementation involves only matrix--vector multiplication, techniques developed both for structured and unstructured sparse matrices may be exploited.

For comparison of computational costs we tested this method with an increasing number of spin particles using different methods to evaluate the exponential.
In particular to examine the error we compared DEC with the {\sl expm} function of \textsl{MATLAB}, that uses a scaling and squaring algorithm with  Pade' approximation.  In this way we evaluate once for all $U=e^{-iL dt}$ where $dt$ is the step size of the simulation, and then at each time--step we propagate $\varrho$:
\begin{equation}\label{prop}
 \varrho_{n+1}=U\varrho_n
\end{equation}
It is well known that in terms of computational costs this simplistic approach performs poorly, so we considered two more realistic model reduction algorithms,  both based on a Krylov subspace expansion.

\subsection{First Alternate Method: Lanczos Iteration} The first one is well known in the literature \cite{saad1,lubich1}.   It evaluates, through a Lanczos algorithm, an orthonormal basis $V_m$ of the Krylov subspace $K_m(L,\varrho_0)$.  

An approximation for $\varrho(t)$ is:
\begin{equation}\label{kryrho}
\varrho(t)=e^{-i L t}\varrho_0 \approx \|\varrho_0\| V_m e^{-i T_m t} e_1, 
\end{equation}
where $T_m$ and $V_m$ come from the Lanczos algorithm. The Lanczos algorithm provides an orthonormal basis set $V_m$ for the Krylov subspace $K_m(L,\varrho_0)$ via a three-term recursion \cite{lancz, cullum}:
\begin{equation}\label{lancz}
 \beta_{j+1}q_{j+1}=Lq_j-\alpha_jq_j-\beta_j q_{j-1} \qquad q_1=\varrho_0.
\end{equation}
$T_m$ is a tridiagonal matrix of size $m \times m$, and $e_1$ is the first vector of the canonical basis of size $n$.
This technique is very powerful for short time simulations, because with few iterations $m$ it is possible to have remarkably good approximations, but for longer times larger Krylov subspaces would be needed to stay close to the real solution. On the other hand if we do not consider enough terms in the Lanczos algorithm for longer times, (\ref{kryrho}) is no longer a reliable approximation. 

It is possible to set a stopping criterion for the Lanczos iterations \cite{lubich1}; for a given $t$ we can find $m$ such that:
\begin{equation}\label{stopcrit}
t [T_m]_{m+1,m} \|e^{-i tT_m}\|_{m,1}\leq \varepsilon.
\end{equation}

To assure a good approximation throughout the whole simulation we set $m$ to verify (\ref{stopcrit}) for $t=\tau$ the total time of the simulation. In this way we have the certainty that the Krylov approximation error is below $\varepsilon$ for  all time $t\leq \tau$.
The equation (\ref{stopcrit}) involves the evaluation of the exponential of a tridiagonal matrix at each step of the Lanczos method; when $m$ is large this operation may become a serious bottleneck for the whole simulation. However in our numerical tests we found out that the order of magnitude of the required $m$ was roughly $m\simeq n/10$, where $n$ is the size of the Liouvillian. 
For this reason although (\ref{stopcrit}) is the proper way to stop the iterations, 
due to the fact that we were more interested in costs comparison than in error analysis, we chose arbitrarily $m=n/10$. 

\subsection{Second Alternate Method: Zero Track Elimination}
The second method considered for comparison is a technique recently developed especially for NMR simulation called ``Zero Track Elimination '', (ZTE) \cite{kuprov2}. This technique is based on the idea of pruning out the elements of $\varrho(t)$ which do not belong to $K(L,\varrho_0)$. In order to reduce the steps needed to evolve the full system, we monitor the elements of  $\varrho(t)$ that stay below a chosen threshold $\xi$ during this first evolution steps and introduce structural zeros based on these observations.
The evolution is then performed in this reduced state space $(\varrho_Z,L_Z)$.

The idea is extremely appealing, as once the propagator for $L_z$ is evaluated all the subsequent steps have the cost of a reduced matrix--vector, and it is possible to use  (\ref{prop}) for the reduced system.
It is claimed \cite{kuprov2} that the error of such an approximation is similar to what would be obtained by considering in the Krylov expansion the contributions coming from high values of $n$ in $L^n\varrho_0$ \cite{kuprov2}.
There are however some drawbacks: 
\begin{itemize}
\item for this method there is no available convergence theory;
\item the performance depends strongly from the initial condition $\varrho_0$, and on $H$. As expected, in our tests the size of the reduced system could change by a factor $2$ depending on the number of interacting spins.
\end{itemize}

\subsection{Summary of Results}

\begin{figure}
\includegraphics[width=0.45\textwidth]{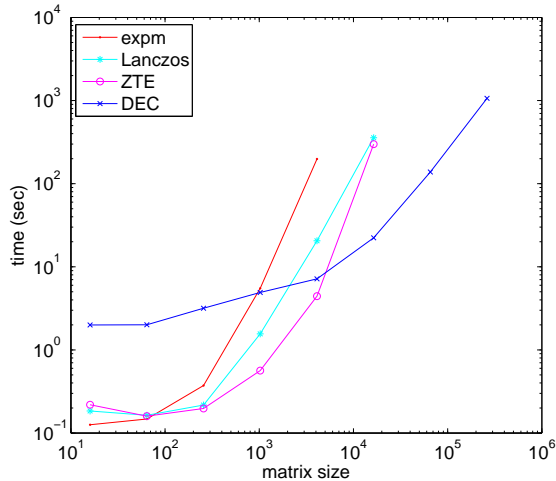}
\caption{Logarithmic comparison of computational costs for Lanczos, Zero Track Pruning (ZTE) and Direct Expectiations via Chebyshev (DEC), with $dt=0.1$ $N=1000$.}
 \label{fig:loglog}
\end{figure}

The error-to-cost (measured in CPU time) diagrams are shown in Figure \ref{fig:loglog} for all the methods described.  It is clear in this example that DEC is more than an order of magnitude more efficient than the alternatives. Obviously especially when the system is over--reduced to slash the computational costs, paying a price in terms of error, DEC still mantains the error below the expected tolerance.    To avoid instabilities coming form the evaluation of the Bessel functions in these numerical tests we set the tolerance to be $\varepsilon = 1e-7$.

\begin{figure}
\includegraphics[width=0.45\textwidth]{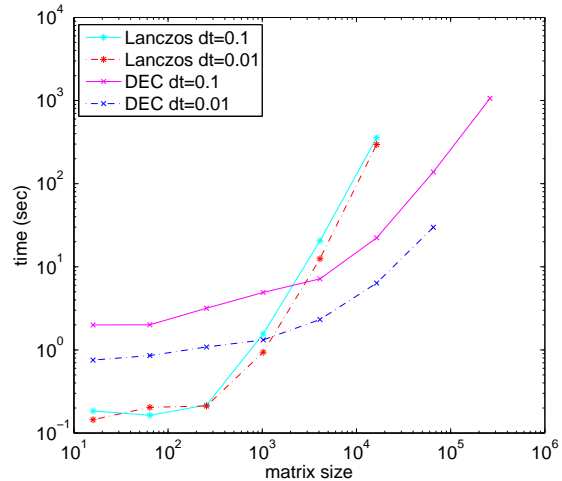}
\caption{Logarithmic comparison of computational costs for Lanczos  and DEC when simulating for the same number of total steps $N$ but with different stepzise $dt$. $N=1000$ in both the cases.}
\label{fig:loglog1}
\end{figure}

DEC performs at its best for short time simulations (i.e. when total time $\tau$ is small), so that we do not need to evaluate a large number of $T_k$, and when at the same time the use of small time step $dt$ is required, as the cost for any step after the first is negligible.
For instance, while for all the other methods the cost of a $1000$ step simulation with $dt=0.1$, is comparable with a simulation of $1000$ steps with $dt=0.1$, for DEC there is a gain of an order of magnitude, see Fig.\ref{fig:loglog1}.

\begin{table}
\begin{ruledtabular}
\caption{Comparison of computational costs,\\ $dt=0.1$, $N=1000$}
\vspace*{0.3cm}
\begin{tabular}{c|c|cc|cc|c}
full size & expm & Lanczos & Reduced\footnotemark[1] & ZTE & Reduced\footnotemark[1] & DEC \\

\hline
16     & 0.12 & 0.15  &  4    & 0.14 &  8      &  1.99    \\
64     & 0.14 & 0.16   &  8    & 0.16   &  24     &  2.00     \\
256    & 0.31  & 0.20   &  25   & 0.20  &  64     &  3.17     \\
1024   & 4.27  & 0.93   &  102  & 0.56  &  160    &  4.91    \\
4096   & 191.4 & 12.66   &  409  & 4.43 &  432    &  7.15   \\
16384  &      & 309.02  &  1638 & 298.57  &  3296   &  22.34    \\ 
65536  &      &        &       &       &         &  138.14    \\ 
262144  &      &        &       &       &         &  1064.7    \\ 
\hline
\end{tabular} 
\label{table:comp1}
\end{ruledtabular}
\footnotemark[1]{Size of $\varrho$ for the reduced system}
\end{table}

\section{Conclusion}
In this article we have presented a new method for simulation of observable in a mixed quantum system. 
By expanding the exponential of the Hamiltonian in Chebyshev polynomials, and exploiting the trace operation performed when evaluating the expectation value of an observable, it is possible to reduce the evolution of any observable to a one--dimension function that can be evaluated directly.

We also presented an optimal algorithm to perform such a calculation, and show how this new method can easily compete in term of computational costs with a variety of model reduction approaches, whilst maintaining the approximation errors below a chosen threshold.

G.M. is very grateful to Arieh Iserles for useful suggestions at the starting of this work.

\appendix*
\section{The DEC algorithm} 
We provide here a detailed description of the algorithm.

\vspace{0.5cm}
\textsl{
{\it inputs}: $L$ hermitian matrix $n \times n$, $\varrho_0$ vector \\
{\it outputs}: expectation value $f(t)$ evaluated at $f(j \Delta t)$, $j=1,\ldots,N$.
\begin{enumerate}
 \item evaluate $\alpha$,$\beta$ via Lanczos s.t. $\alpha=\min(\lambda_i)$, $\beta=\max(\lambda_i)$;
\item scale $L$ and get $L_s$
\item evaluate $T_0= {\rm Id}\varrho_0$, $T_1=L_s\varrho_0$
\item {\bf while} $\|c_k\|< \varepsilon$
\item \hspace{1cm} $c_k=(2-\delta_{k,0})(-i)^k J_k(\tau S)$, $\tau=$ total time 
\item \hspace{1cm} $T_{k+1}=2T_kL_s-T_{k-1}$
\item \hspace{1cm} store $\tilde{T}_k={\rm Trace}\{(T_k\varrho_0)\hat{Q}\}$
\item {\bf end}
\item {\bf for} $j=1:{\rm N}$
\item \hspace{1cm} re-evaluate the $c_k$ at different time $t=j d t$ 
\item \hspace{1cm} $f(j)=\sum c_k \tilde{T}_k$
\item {\bf end}
\end{enumerate}
}
\bibliography{plain}

\begin{thebibliography}{9}
\bibitem{nineteen} C. Moler and C. Van Loan, SIAM Review, {\bf 45}(1), 3--49  (2003).
\bibitem{saad1} Y. Saad, SIAM J. Num. Anal., {\bf 29}(1), 209--228 (1992).

\bibitem{lubich1} M. Hochbruck and C. Lubich, SIAM J. Num. Anal. {\bf 34}(5), 1911--1925 (1997).
\bibitem{schulze} J.C. Schulze, P.J. Schmid and J.L. Sesterhenn, Int. J. Numer. Meth. Fluids, {\bf 60}, 591--609 (2009).

\bibitem{lancz} C. Lanczos,  J. Res. Nat. Bur. Standards, {\bf 45}, 255--282 (1950).
\bibitem{golubok} G.H. Golub and C.F. Van Loan, \textit{Matrix Computations}, (The John Hopkins University Press, Baltimore, 1996).
\bibitem{cullum} J. Cullum and R.A. Willoughby, \textit{Lanczos Algorithms for large symmetric eigenvalue computations}, (Birkh\"{a}user, Boston, 1985).
\bibitem{tal} H. Tal--Ezer and R. Kosloff, J. Chem. Phys. {\bf 81}, 3967--3971 (1984).
\bibitem{lubichbok} C. Lubich \textit{From Quantum to Classical Molecular Dynamics: Reduced Models and Numerical Analysis}, (Zurich Lectures in Advanced Mathematics, Zurich, 2008).

\bibitem{saad_lanc} Y. Zhou, Y. Saad, M.L. Tiago and J.R. Chelikowsky, J. Comp. Phys., {\bf 219}(1), 172--184, (2006).
\bibitem{handbok} M. Abramowitz and  I.A. Stegun \textit{Handbook of mathematical functions}, (Dover, New York, 1964).
\bibitem{backward} F.W.J. Olver and D.J. Sookne, Math. of Comp., {\bf 26}(120), 941--947, (1972).

\bibitem{Levitt} M.H. Levitt, \textit{ Spin Dynamics}, (Wiley, Chichester, 2001).
\bibitem{kuprov} I. Kuprov, J. Magn. Reson., {\bf 195}, 45--51, (2008).
\bibitem{simpson} M. Bak, J.T. Rasmussen and N.C. Nielsen, J. Magn. Reson. {\bf 147}, 296--330, (2000).
\bibitem{spinev} M.Veshtort, R.G.Griffin, J. Magn. Reson., {\bf 178}, 248--282, (2006).

\bibitem{kuprov2} I. Kuprov, J. Magn. Reson., { \bf 195},  45--51, (2008).

\end{thebibliography}

\end{document}